\documentclass[aip,apl,reprint,twocolumn]{revtex4-1}
\usepackage[dvipdfmx]{graphicx}
\usepackage[version=3]{mhchem}
\usepackage{braket}
\usepackage{natbib}
\usepackage{here}
\usepackage{dcolumn}
\usepackage{bm}
\newcommand{\LSFO}{\ce{La1/3Sr2/3FeO3}}
\begin{document}
\title{Thickness dependence and dimensionality effects of charge and magnetic orderings in \ce{La1/3Sr2/3FeO3} thin films}
\author{K.~Yamamoto}
\homepage{http://sites.google.com/site/yamakolux/}
\email{yamako@issp.u-tokyo.ac.jp}
\affiliation{Institute for Solid State Physics, University of Tokyo, Kashiwanoha, Chiba 277-8581, Japan}
\affiliation{Department of Physics, University of Tokyo, Hongo, Tokyo 113-0033, Japan}
\author{Y.~Hirata}
\affiliation{Institute for Solid State Physics, University of Tokyo, Kashiwanoha, Chiba 277-8581, Japan}
\affiliation{Department of Physics, University of Tokyo, Hongo, Tokyo 113-0033, Japan}
\author{M.~Horio}
\affiliation{Department of Physics, University of Tokyo, Hongo, Tokyo 113-0033, Japan}
\author{Y.~Yokoyama}
\affiliation{Institute for Solid State Physics, University of Tokyo, Kashiwanoha, Chiba 277-8581, Japan}
\affiliation{Department of Physics, University of Tokyo, Hongo, Tokyo 113-0033, Japan}
\author{K.~Takubo}
\affiliation{Institute for Solid State Physics, University of Tokyo, Kashiwanoha, Chiba 277-8581, Japan}
\author{M.~Minohara}
\affiliation{Institute of Materials Structure Science, High Energy Accelerator Research Organization, Tsukuba, Ibaraki 305-0801, Japan}
\author{H.~Kumigashira}
\affiliation{Institute of Materials Structure Science, High Energy Accelerator Research Organization, Tsukuba, Ibaraki 305-0801, Japan}
\author{Y.~Yamasaki}
\affiliation{Department of Applied Physics and Quantum-Phase Electronics Center, University of Tokyo, Hongo, Tokyo 113-8656, Japan}
\affiliation{RIKEN Center for Emergent Matter Science, Wako, Saitama 351-0198, Japan}
\author{H.~Nakao}
\affiliation{Institute of Materials Structure Science, High Energy Accelerator Research Organization, Tsukuba, Ibaraki 305-0801, Japan}
\author{Y.~Murakami}
\affiliation{Institute of Materials Structure Science, High Energy Accelerator Research Organization, Tsukuba, Ibaraki 305-0801, Japan}
\author{A.~Fujimori}
\affiliation{Department of Physics, University of Tokyo, Hongo, Tokyo 113-0033, Japan}
\author{H.~Wadati}
\affiliation{Institute for Solid State Physics, University of Tokyo, Kashiwanoha, Chiba 277-8581, Japan}
\affiliation{Department of Physics, University of Tokyo, Hongo, Tokyo 113-0033, Japan}
\date{\today}
\begin{abstract}
We investigate the thickness effects on charge and magnetic orderings in Fe perovskite oxide \ce{La_{1/3}Sr_{2/3}FeO3}/\ce{SrTiO3} thin films by hard x-ray and resonant soft x-ray scattering (RSXS) with changing thin film thickness systematically.
We found that the correlation lengths of the magnetic ordering along the in-plane and out-of-plane directions are comparable and proportional to the thickness, and shows stronger thickness dependence than those of charge orderg.
%the thickness dependence of correlation length of charge ordering is smaller than that of magnetic orderings.
The magnetic ordered states disappear when the correlation length of magnetic ordering decreases to that of charge ordering through the intrinsic thickness effects.
Surface sensitive grazing-incident RSXS revealed that the orderings exist even in the surface region, which indicates that the observed orderings is not affected by surface effect like oxygen vacancies.
Critical thickness is in 5-15 nm, which corresponds to 4-11 antiferromagnetic ordering period.
This critical value seems to be common to other ferromagnetic oxide thin films.
\end{abstract}
\maketitle

%\section{Introduction}
Charge, spin, and orbital orderings in 3\textit{d} transition metal perovskite oxides closely relate to the origin of remarkable physics phenomena such as metal-insulator transition, superconductivity and giant magnetoresistance.\cite{Imada1998,Tokura2000} 
For instance, stripe structures of spin and charge ordering in superconducting cuprates were reported~\cite{Tranquada1995,Tranquada1998} and the charge density wave was found to compete with superconducting state.\cite{Ghiringhelli2012,Blanco-Canosa2013,He2016}

\begin{figure}
\begin{center}
  \includegraphics[clip,width=9cm]{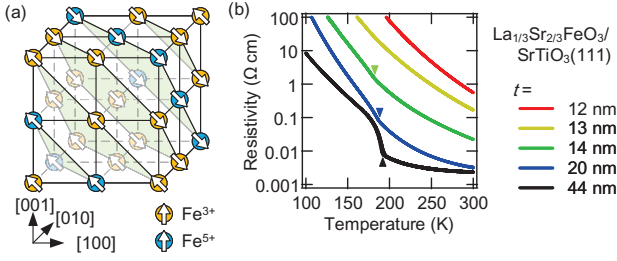}
  \caption{Charge disproportionation in \LSFO. (a) Schematic drawing of charge and spin ordered structure of \LSFO\ determined by neutron~\cite{Battle1990} and electron diffraction.\cite{Li1997} 
(b) Temperature dependence of electric resistivity of \ce{La_{1/3}Sr_{2/3}FeO3}/\ce{SrTiO3} thin films reproduced from \citeauthor{Minohara2016}.~\cite{Minohara2016}}
\label{fig:intro}
  \end{center}
\end{figure}

One of such 3\textit{d} transition-metal oxides, \LSFO\ attracts much attention due to its unusual high valence state and charge disproportionation.
M\"ossbauer spectroscopy showed charge disproportionation of 3\ce{Fe^{3.67+}}$\rightarrow$2\ce{Fe^{3+}}+\ce{Fe^{5+}} accompanied by an antiferromagnetic ordering at critical temperature $T_\mathrm{CD}\approx 190~\mathrm{K}$.\cite{Takano1983,Battle1988}
Electronic conductivity shows a transition associated with the charge disproportionation transition at $T_\mathrm{CD}$.\cite{Takano1983}
Neutron scattering study revealed that charge and magnetic orderings exist along the [111] direction and Fe layers stack in the sequence of \ce{Fe^{3+} ^ -Fe^{5+} ^ -Fe^{3+} ^ -Fe^{3+} v -Fe^{5+} v -Fe^{3+} v } as shown in Fig.~\ref{fig:intro}(a).\cite{Battle1990}
Electron diffraction demonstrated structural distortion along the $[111]$ direction.\cite{Li1997}
This stacking layer structure is also supported by the exchange interaction determined by inelastic neutron scattering~\cite{McQueeney2007} and Hartree-Fock calculation.\cite{Mizokawa1998}
As for the electronic structure, x-ray absorption spectroscopy of Fe 2{\it p} $\rightarrow$ 3{\it d} and O 1{\it s} $\rightarrow$ 2{\it p} suggests that holes induced by Sr doping exhibits O 2{\it p} hole character.\cite{Abbate1992}
The unusually high valence states \ce{Fe^{4+}} and \ce{Fe^{5+}} are expressed rather as \ce{3{\it d}^5\underline{\it L}} and \ce{3{\it d}^5\underline{\textit{L}}^2}, where \ce{\underline{\it L}} represents an O 2{\it p} hole.\cite{Abbate1992,Bocquet1992,Imada1998}

The charge ordering exists even in thin films.\cite{Wadati2005,Okamoto2010,Ueno2006,Sichel-Tissot2013,Xie2014}
This is considered to be due to the small lattice distortion accompanied by the charge disproportionation in contrast to the case of \ce{Fe^{4+}} oxide \ce{CaFeO3}~\cite{Matsuno2004} or Mn perovskite oxides such as $\mathrm{Pr}_{1-x}\mathrm{Ca}_x\mathrm{MnO}_3$ and $\mathrm{Nd}_{1-x}\mathrm{Sr}_x\mathrm{MnO}_3$.\cite{Tokura2006}
For this reason, \LSFO\ thin film is a suitable material for extracting intrinsic effects of thickness and dimensionality on electronic states without the influence of lattice distortion.
However, there have been no studies of thickness effect except for resistivity measurements.\cite{Minohara2016,Devlin2014}
\citeauthor{Minohara2016} studied the precise thickness dependence of resistivity in \LSFO/\ce{SrTiO3} thin films and found that the ordered states were suppressed with thickness $t$ below 14 nm as shown by $\rho-T$ curves in Fig.~\ref{fig:intro}(b).\cite{Minohara2016} 
Resistivity becomes too high to determine the resistivity jump with decreasing thickness due to the insulating \ce{SrTiO3} substrates and resistivity is not a direct evidence for the orderings.
Therefore, x-ray is a suitable technique for obtaining a detailed view of and determining the existence of the orderings in \LSFO thin films.

\citeauthor{Okamoto2010} performed hard x-ray scattering for charge ordering and resonant soft x-ray scattering (RSXS) for magnetic ordering at the Fe $2p_{3/2}\rightarrow3d$ absorption.\cite{Okamoto2010}
One cannot observe the magnetic ordering peak $\left(\frac{1}{6}\frac{1}{6}\frac{1}{6}\right)$ by hard x-ray~\cite{Okamoto2010} due to the small cross section of magnetic scattering.\cite{Blume1985}
%\LSFO\ is an antiferromagnetic material and magnetization measurement cannot be used for observing antiferromagnetic orderings.
RSXS is a suitable technique to study magnetic orderings in thin films and small crystals.\cite{Blume1985,Okamoto2010,Fink2013,Matsuda2015,Wadati2012,Huang2006,Wadati2014,Ghiringhelli2012,Blanco-Canosa2013,He2016,Yamasaki2016}
We obtained a strong sensitivity of magnetic scattering at the Fe $2p\rightarrow 3d$ absorption edge due to strong spin-orbit coupling of Fe $2p$ core levels.
In addition to reflection geometry RSXS, we performed grazing-incident RSXS (GI-RSXS).
When the incident angle is below the total reflection angle, the probing depth of x-ray becomes small and one can access surface-sensitive information.
This technique was used to study surface states of ordering and revealed that surface melting in \ce{La_{0.5}Sr_{1.5}MnO_4}~\cite{Wakabayashi2007,Wilkins2011} and surface orbital ordering in \ce{LaCoO3} thin film.\cite{Yamasaki2016}

\begin{figure}[h]
  \begin{center}
    \includegraphics[clip,width=8cm]{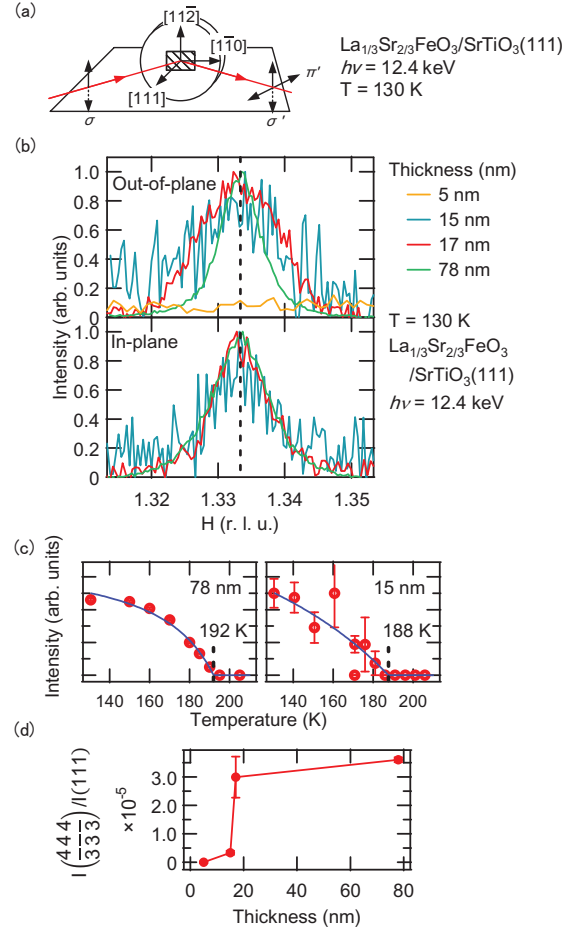}
    \caption{Observation of charge ordering by hard x-ray diffraction.
      (a) Experimental setup. $\pi$, $\pi'$ and $\sigma'$ indicate the directions of polarizations of x-rays.
      (b) Charge ordering peaks $\left(\frac{4}{3}\frac{4}{3}\frac{4}{3}\right)$ observed along the out-of-plane and in-plane directions at 130 K.
      %The peaks of thin films with thickness of 15, 17, 78 nm are normalized to the peak height and the peak of thin film with a thickness of 5 nm is multiplied by the same value of 17 nm.
      %Dashed lines indicate the position of $\left(\frac{4}{3}\frac{4}{3}\frac{4}{3}\right)$
      (c) Temperature dependence of the charge ordering peak and the area intensity. The lines are guides to the eye.
(d) Charge ordering peak intensity normalized by the lattice Bragg peak (111) intensity plotted as a function of thickness.}
    \label{fig:hx}
  \end{center}
\end{figure}

\begin{figure}[h]
\begin{center}
  \includegraphics[clip,width=8cm]{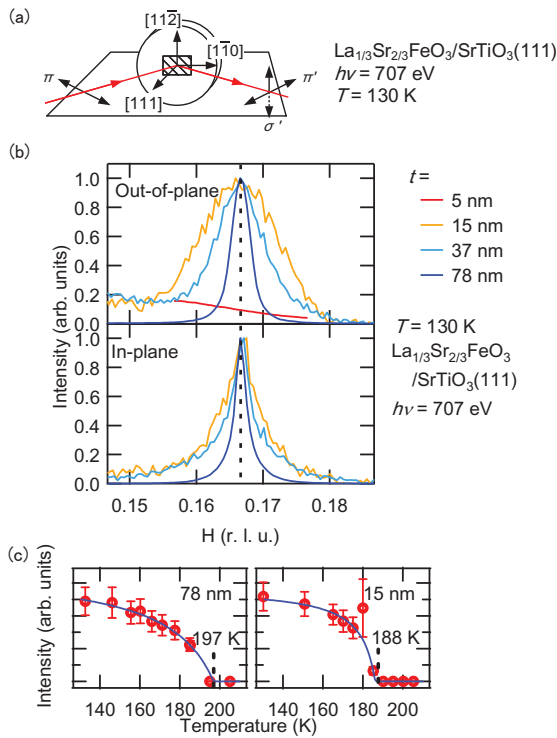}
  \caption{Observation of magnetic ordering by RSXS.
 (a) Experimental setup. $\pi$, $\pi'$ and $\sigma'$ indicate the directions of polarizations of x-rays.
    (b) Magnetic ordering peaks $\left(\frac{1}{6}\frac{1}{6}\frac{1}{6}\right)$ observed along the out-of-plane and in-plane directions at 130 K.
    %The peaks of thin films with thickness of 15, 37, 78 nm are normalized to the peak height.% and the peak of thin film with a thickness of 5 nm is normalized to the value of $H = 0.157$.
    %Dashed lines indicate the position of $\left(\frac{1}{6}\frac{1}{6}\frac{1}{6}\right)$.
 (c) Temperature dependence of the magnetic ordering peak and the peak intensity. The lines are guides to the eye. }
  \label{fig:sx}
  \end{center}
\end{figure}

\begin{figure}
  \begin{center}
    \includegraphics[clip,width=7cm]{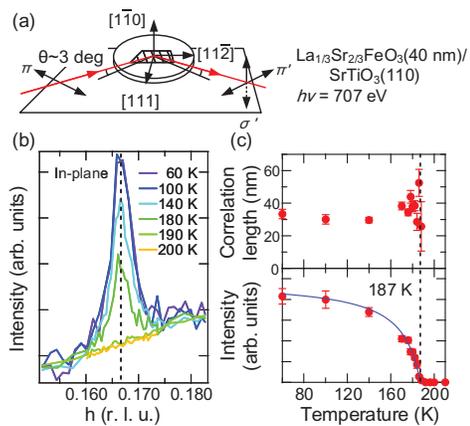}
    \caption{Observation of the magnetic ordering peak by GI-RSXS.
 (a) Experimental setup. $\pi$, $\pi'$ and $\sigma'$ indicate the directions of polarizations of x-rays.
 (b) Magnetic ordering peaks $\left(\frac{1}{6}\frac{1}{6}\frac{1}{6}\right)$.
 (c) Correlation length and peak area intensity plotted as a function of temperature.}
    \label{fig:gi}
  \end{center}
\end{figure}

%\section{Experiment}
The \LSFO\ thin films were grown on \ce{SrTiO3}(111) and (110) substrates by pulsed laser deposition method. 
The details of the sample fabrication were described elsewhere.\cite{Minohara2016}
The thickness of the thin films was determined from the Laue fringes around the (111) Bragg peaks (not shown).
We performed hard x-ray scattering experiments for \ce{La_{1/3}Sr_{2/3}FeO3(t = 5,\, 15,\, 17\, and\, 78\, nm)/SrTiO3(111)} thin films at BL-4C of Photon Factory, KEK.
We measured the charge ordering peak $\left(\frac{4}{3}\frac{4}{3}\frac{4}{3}\right)$ with hard x-ray.
The experimental geometry is shown in Fig.~\ref{fig:hx}(a).
The incident photon energy of x-ray was 12.4 keV.
Incident x-ray was $\sigma$-polarized and scattered x-ray was detected without polarization analysis and hence includes $\sigma'$- and $\pi'$-polarized photons.
We performed RSXS experiments for \ce{La_{1/3}Sr_{2/3}FeO3(t = 5,\, 17,\, 37\, and\, 78\, nm)/SrTiO3(111)} thin films at the BL-19B of Photon Factory.
We observed the magnetic ordering scattering peak  $\left(\frac{1}{6}\frac{1}{6}\frac{1}{6}\right)$ by RSXS.
The incident photon energy was 707 eV at the Fe $2p_{3/2}\rightarrow 3d$ absorption edge.
Incident x-ray was $\pi$-polarized and scattered photon was detected by an energy-resolved silicon drift detector without polarization analysis.
We removed O $\mathrm{K}_\alpha$ fluorscence emission by tuning energy windows to the Fe L edge.
The geometry is shown in Fig.~\ref{fig:sx}(a).
The sample was \ce{La_{1/3}Sr_{2/3}FeO3(40\, nm)/SrTiO3(110)}, which has in-plane modulation vectors.
In order to investigate the surface states, we performed GI-RSXS in the geometry shown in Fig.~\ref{fig:gi}(a) using the sample \ce{La_{1/3}Sr_{2/3}FeO3(40\, nm)/SrTiO3(110)}, which has in-plane modulation vectors.

%\section{Results and Discussion}

\begin{figure}
  \begin{center}
    \includegraphics[clip,width=8cm]{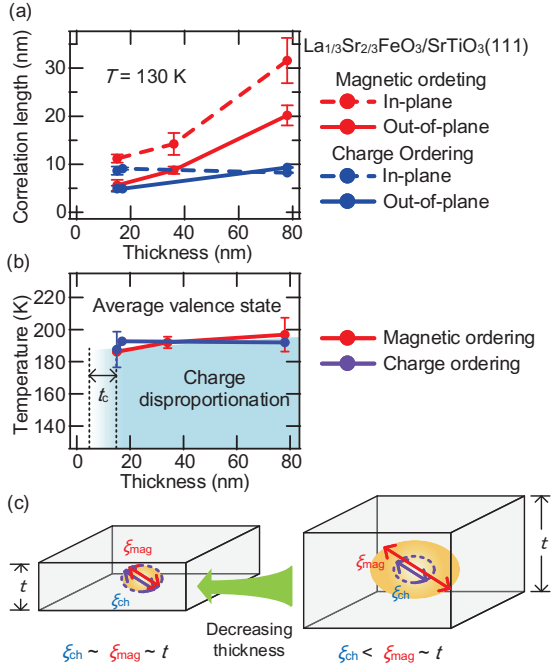}
    \caption{Thickness dependence in \LSFO thin films revealed by x-ray scattering.
      (a) Correlation length of magnetic and charge ordering for various thickness thin films. These values were determined by hard and soft x-ray scattering results.
      (b) Phase diagrams estimated from the temperature dependence of charge and magnetic ordering peak intensities in \LSFO/\ce{SrTiO3}(111) thin films. $t_{\rm c}$ indicates the critical thickness of the charge disproportionation.
      (c) Schematic illustration of the thickness dependent change of ordered states.}
    \label{fig:tccorrl}
 \end{center}
\end{figure}

Figure~\ref{fig:hx}(b) shows the charge ordering peak $\left(\frac{4}{3}\frac{4}{3}\frac{4}{3}\right)$ along the out-of-plane $\left[111\right]$ direction and along the in-plane $\left[1\overline{1}0\right]$ direction.
The charge orderings exist in $t =\ 15\ \mathrm{nm}$, and disappears in $t =\ 5\ \mathrm{nm}$.
The peak width shows little dependence on thickness.
%Thickness dependence of peak width is small and the charge orderins disappear in $t=5\ \mathrm{nm}$.
Figure~\ref{fig:hx}(c) shows the temperature dependence of the peak intensity. The critical temperature $T_\mathrm{CD}$ was determined from this dependence. 
Figure~\ref{fig:hx}(d) shows that the charge ordering peak intensity drops sharply when the thickness is reduced to 15 nm.
This drop suggests that critical thickness $t_\mathrm{c}$ exists and the value is $t_{c}\sim10\ \mathrm{nm}$, which is in good agreement with the previous electric resistivity study.\cite{Minohara2016}
Figure~\ref{fig:sx}(b) shows the magnetic ordering peak $\left(\frac{1}{6}\frac{1}{6}\frac{1}{6}\right)$ along both directions.
The peak width increases when thickness becomes thinner and the peak disappears at $t=5\ \mathrm{nm}$.
This disappearance is also consistent with the charge ordering behavior shown in Fig.~\ref{fig:hx}(b, d).
The correlation length is given as the inverse of the half width of half maximum, namely, $1/\Delta q$.
This peak width tendency of the magnetic ordering indicates that correlation length decreases when thickness becomes thinner.
The critical temperature was also determined by the temperature dependence of the peak intensity as shown in Fig.~\ref{fig:sx}(c).
The correlation length and the critical temperature of the charge and magnetic ordered states are summarized in Fig.~\ref{fig:tccorrl} and will be discussed later.

It may be the case that surface states differ from bulk states by possible oxygen vacancies due to the high valence state of \ce{Fe^{3.67+}}.
Effects of oxygen vacancy affects the orderings~\cite{Sichel-Tissot2013,Xie2014} and probing surface state is important for studying the thickness dependence.
We performed GI-RSXS study in order to investigate the surface states.
We studied \ce{La_{1/3}Sr_{2/3}FeO3}(40 nm)/\ce{SrTiO3}(110) thin film with the in-plane modulation vector and measured the magnetic orderings peak $\left(\frac{1}{6}\frac{1}{6}\frac{1}{6}\right)$.
Its probing depth was determined to be approximately 2 nm by calculation from x-ray absorption spectroscopy and the incident angle.
Figure \ref{fig:gi}(b) shows the result of GI-RSXS and Fig.~\ref{fig:gi}(c) shows the correlation length and the transition temperature.
$T_\mathrm{CD}$ and the peak position of surface states are the same as those of bulk states shown in Fig.~\ref{fig:tccorrl} and correlation length is in the same order of the results of the bulk state.
These results indicate that the surface states of \LSFO\ thin films are the same as the bulk state and the measured thickness dependence is not due to the surface effects but due to intrinsic dimensionality effects.

The obtained correlation length is plotted as a function of thickness in Fig.~\ref{fig:tccorrl}(a) at 130 K.
Out-of-plane correlation length of magnetic orderings is proportional to the thickness.
Geometrical limitation of thickness explains this behavior of the out-of-plane correlation length.
However, the in-plane correlation length also decreases when the thickness of the thin film becomes thinner.
This behavior suggests that magnetic orderings have an isotropic correlation.
Comparing with correlation length of magnetic ordering~$\xi_{\rm mag}=1/\Delta q(\frac{1}{6}\frac{1}{6}\frac{1}{6})$, the thickness dependence of the correlation length of charge orderings~$\xi_{\rm ch}=1/\Delta q(\frac{4}{3}\frac{4}{3}\frac{4}{3})$ is much weaker than that of magnetic orderings. 
When the thickness is sufficiently large ($\sim78\ \mathrm{nm}$), the correlation length of magnetic ordering is in the same order of thickness ($\xi_\mathrm{ch}<\xi_\mathrm{mag}\sim t$).
When the thickness and $\xi_\mathrm{mag}$ reach $\xi_\mathrm{ch}$ at the thickness of approximately 15 nm ($\xi_\mathrm{ch}\sim\xi_\mathrm{mag}\sim t$), the orderings disappear.
This indicates that a minimal value of domain size of $\sim 15~\mathrm{nm}$ exists and thickness limitation makes the correlation length of magnetic ordering below the minimal value and suppresses the orderings.
This behavior is schematically shown in Fig.~\ref{fig:tccorrl}(c).
The obtained critical thickness of 5-15 nm corresponds to 4-11 units with the antiferromagnetic ordering period of 6 u. c. as a unit.
This critical number of units seems to be common to other ferromagnetic perovskite oxide $\mathrm{La}_{1-x}\mathrm{Sr}_x\mathrm{MnO}_3(x=0.3-0.4)$ (3-8 units~\cite{Huijben2008,Yoshimatsu2009}).
We obtained the critical temperature $T_\mathrm{CD}$ from temperature dependence of scattering peak area intensity and it is plotted as a function of thickness as shown in Fig.~\ref{fig:tccorrl}(b).
$T_\mathrm{CD}$ slightly decreased when thin film became thinner and $T_\mathrm{CD}$ of charge and magnetic orderings were almost the same.
This slight decrease is consistent with other magnetic ordered compounds and matches the result of resistivity study.\cite{Minohara2016}

%\section{Conclusion}
In summary, we performed hard x-ray scattering and RSXS experimets for the charge and magnetic orderings in \LSFO/\ce{SrTiO3}~thin films with changing the thickness systematically.
We also performed GI-RSXS experiment in order to investigate surface states of \LSFO\ thin films with the unusually high valence state.
Magnetic ordering is stable even in the surface region in spite of its high valence states and this suggests that the thickness dependence comes from the intrinsic geometrical effects.%this is desirable nature from the point of view of application to electric devices.
The correlation length of the magnetic ordering $\xi_\mathrm{mag}$ is comparable and proportional to the thickness.
Reduced thickness suppresses the ordered states at the critical temperature $t_\mathrm{C}$ through this relationship between $t$ and $\xi_{mag}$.
It will be interesting to study systematically thickness dependences of charge, magnetic or orbital orderings with x-ray scattering for various thin films and elucidate behaviors especially around critical thicknesses.

\begin{acknowledgments}
X-ray scattering measurements were performed under the approval of the Photon Factory Program Advisory Committee (Proposals No. 2015G556, 2016PF-BL-19B, 2015S2-007) at the Institute of Material Structure Science, KEK. 
This work was partially supported by the Ministry of Education, Culture, Sports, Science and Technology of Japan (X-ray Free Electron Laser Priority Strategy Program).
K. Y. acknowledges the support from ALPS program of the University of Tokyo.
\end{acknowledgments}

\end{document}